\documentclass[journal]{vgtc} % final (journal style)
\usepackage{cite}
\usepackage[bottom]{footmisc}
\usepackage{mathptmx}
\usepackage{graphicx}
\usepackage{siunitx}
\usepackage{times}
\usepackage[T1]{fontenc}
\usepackage{enumitem}
\usepackage[table]{xcolor} % loads also »colortbl«
\usepackage{wrapfig}
\usepackage{multirow}
\usepackage{adjustbox}
\usepackage{mwe}
\usepackage{graphbox}
\usepackage{dirtytalk}
\usepackage{soul}
\usepackage{mdframed}
\usepackage{tabularx}
\usepackage{xcolor}
\usepackage{enumitem}
\definecolor{CindySalmon}{RGB}{232, 125, 114}
\definecolor{greenB}{RGB}{77, 175, 74}
\definecolor{purpleF}{RGB}{152,78,163}

\usepackage[bookmarks,backref=true,linkcolor=black]{hyperref} %,colorlinks
\hypersetup{
 pdfauthor = {},
 pdftitle = {},
 pdfsubject = {},
 pdfkeywords = {},
 colorlinks=true,
 linkcolor= black,
 citecolor= black,
 pageanchor=true,
 urlcolor = black,
 plainpages = false,
 linktocpage
}

\onlineid{1151} 
\vgtccategory{Theoretical and Empirical} % declare the paper category, only shown in review mode
\vgtcinsertpkg % allow for this line if you want the electronic option to work
\title{Seeing What You Believe or Believing What You See? \\Belief Biases Correlation Estimation
}

\author{Cindy Xiong, Chase Stokes, Yea-Seul Kim, Steven Franconeri}
\authorfooter{ % insert punctuation at end of each item
\item
Cindy Xiong is with UMass Amherst.
\newline
E-mail: cindy.xiong@cs.umass.edu.
\item
Chase Stokes is with UC Berkeley.
\item
Yea-Seul Kim is with the University of Wisconsin Madison,
\item
Steven Franconeri is with Northwestern University
}

\shortauthortitle{Xiong \MakeLowercase{\textit{et al.}}: }

\abstract{When an analyst or scientist has a belief about how the world works, their thinking can be biased in favor of that belief. Therefore, one bedrock principle of science is to minimize that bias by testing the predictions of one’s belief against objective data. But interpreting visualized data is a complex perceptual and cognitive process. Through two crowdsourced experiments, we demonstrate that supposedly objective assessments of the strength of a correlational relationship can be influenced by how strongly a viewer believes in the existence of that relationship. Participants viewed scatterplots depicting a relationship between meaningful variable pairs (e.g., number of environmental regulations and air quality) and estimated their correlations. They also estimated the correlation of the same scatterplots labeled instead with generic 'X' and 'Y' axes. In a separate section, they also reported how strongly they believed there to be a correlation between the meaningful variable pairs. Participants estimated correlations more accurately when they viewed scatterplots labeled with generic axes compared to scatterplots labeled with meaningful variable pairs. Furthermore, when viewers believed that two variables should have a strong relationship, they overestimated correlations between those variables by an r-value of about 0.1. When they believed that the variables should be unrelated, they underestimated the correlations by an r-value of about 0.1. While data visualizations are typically thought to present objective truths to the viewer, these results suggest that existing personal beliefs can bias even objective statistical values people extract from data.}


\begin{document}
\maketitle

\section{Introduction}
\begin{figure}[h!]
\centering
\includegraphics[width = \columnwidth]{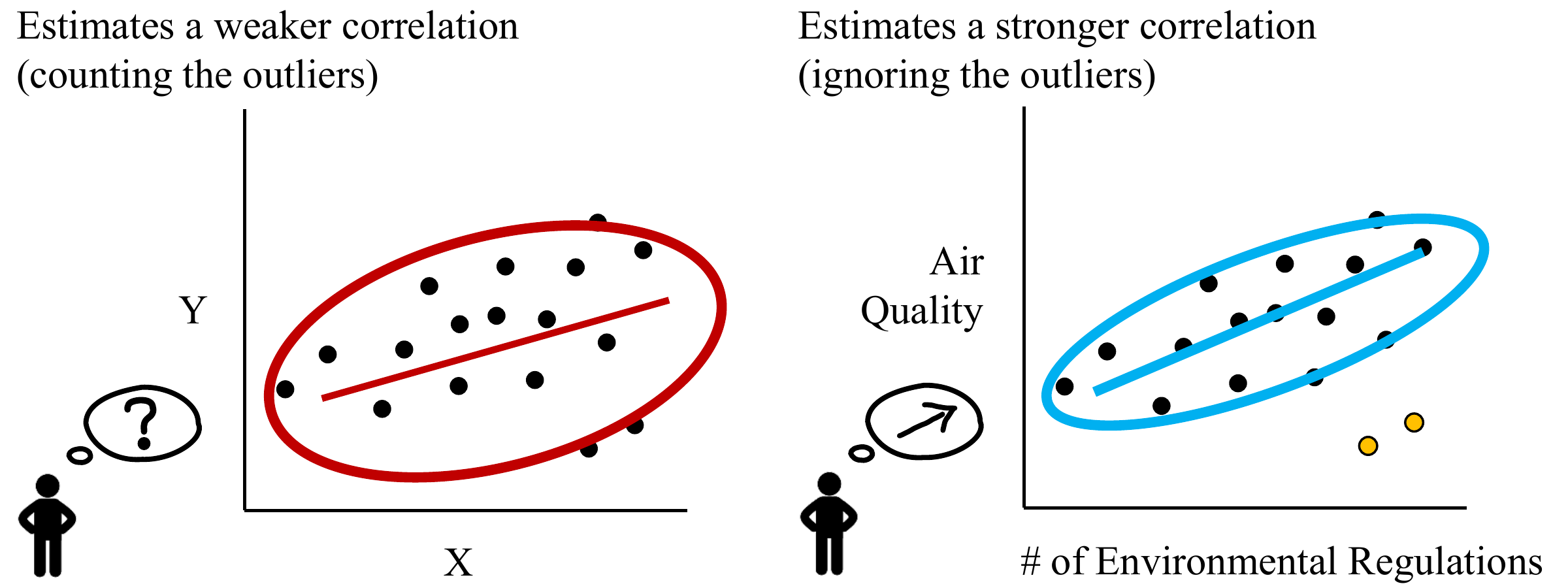}
 \caption{People might induce themselves to see a correlation as weak or strong, depending on existing beliefs about data.}
 \label{fig:motivatingExample}
\end{figure}

Your new data have arrived. With the intent of showing a relationship between the strength of environmental regulations in a region and that region's air quality, you plot your data values in a scatterplot. At first, the correlation is barely noticeable. But you then notice a small set of outliers. If you ignore those, the correlation looks as strong as you expected, as illustrated in Figure \ref{fig:motivatingExample}.

% Psych example of motivated perception
This hypothetical example may be more common than one would hope. This type of data analysis is intended to proceed in a purely `bottom-up' fashion, driven only by the data itself. However, motivation, belief, and experience can bias us across a complex set of `top-down' interactions \cite{lupyan2017changing, luo2019motivated}. When people viewed an ambiguous figure which was carefully balanced to be interpreted either as the letter B or the number 13, those who expected letters were more likely to see `B', while those who expected to see numbers were more likely to see `$1\!3$' \cite{balcetis2006}. 

% data Vis example of motivated perception
Similar top-down biases also seem to appear in data analysis.
When viewing a visualization depicting global temperature trends, liberals focused more on the increasing trend, confirming their worries about climate change. Conservatives focused more on the overall flatness of the line and the small effect size of the increase, which were more consistent with doubts about those concerns \cite{luo2019motivated}.

In another study, participants were told a `top-down' story explaining a pattern present across randomly chosen data values in a visualization and were then explicitly told to ignore that story. Yet, later ratings of the `bottom-up' salience of those values were higher for that subset of values than for others. The same participants even predicted that \textit{other} people, naive to the top-down story, would also find those values more salient \cite{xiong2019curse}.

% Study Motivation
These past studies of top-down influences in data visualizations focus on how top-down information can bias the visual salience of certain values or attributes. In contrast, the present studies provide a quantitative estimate of how powerful this type of influence can be, providing a measurement of how a belief can change an `objective' correlation estimate. 
We use scatterplots to show those correlations, as those plots are among the most commonly used visualizations and are especially valuable for depicting correlation \cite{rensink2010perception, harrison2014ranking, kim2018assessing}, which is a critical attribute that impacts data-driven decisions \cite{friendly2005early}. 
Additionally, the perception of correlation has been relatively well-studied. 
People are fairly accurate at determining the Pearson's r correlation in scatterplots \cite{rensink2010perception}, and recent work has begun to explore the underlying visual features that people use as proxies when estimating correlations (e.g., the shape of the ellipse that encloses the scattered dots) \cite{yang2018correlation}.%, and we take a different approach to additionally investigate the non-visual factors that people use to estimate correlations, such as their existing belief. 
\newline
\vspace{-2mm}

% Research Questions and Hypothesis
\noindent \textbf{Hypothesis:} We hypothesize that people 
% will estimate a stronger correlation when they have a predisposed strong belief in the relationship that the scatterplot depicts and a weaker correlation when they have a weak belief.
will estimate correlations more accurately when they see scatterplots labeled with a neutral pair of variables that are not associated with existing beliefs or abstracted data labels (e.g., X vs. Y). 
We also hypothesize that the amount of deviation in people's correlation estimation would be associated with the strength of their belief, such that they would report a higher correlation for the same scatterplot when it depicts something that they strongly believe to be correlated and a lower correlation when the plot depicts something that they do not strongly believe to be correlated.

% Avoiding the Perception vs. Cognition Debate
We are careful to state these hypotheses as beliefs affecting what people \textit{report}, as opposed to them affecting what people \textit{see}. Beliefs might affect various stages of processing on the road to a correlation estimate, including what people perceive, what values they preferentially attend, or what values they weigh most heavily when deciding what numeric estimate to report. 
Depending on a researcher's perspective, these stages can have boundaries that are firm, as some have argued that top-down beliefs and motivations do not directly impact perception \cite{firestone2016cognition}, or on a continuum, as others have argued for a `cognitive penetrability of perception' suggesting that top-down effects can directly influence what we see \cite{vetter2014varieties, raftopoulos2001perception}.
\newline
\vspace{-2mm}

% Key Findings/Summary of Experiments
\noindent \textbf{Experiment Overview:} We present two empirical experiments in which participants estimated correlations for identical data sets in scatterplots.
The scatterplots have axes either labeled with meaningful variable pairs (e.g., environmental regulation and air quality) or labeled with abstracted variables `X' and `Y' (separated by a distractor task). 
In a separate section, participants reported how strongly they believed there to be a correlation between the variable pairs, either after they saw the scatterplots (Experiment 1a) or before (Experiment 2). 
In both studies, we found that people who believed the meaningful variable pairs to be correlated overestimated their correlation values significantly, while people who did not hold those beliefs underestimated the correlations significantly, in comparison to the scatterplots labeled with abstracted variables. 
We additionally conducted a control experiment (Experiment 1b) to test whether the differences in correlation estimate were due to repeated exposure to the same scatterplots (i.e., having seen the same dataset in a scatterplot twice, one with meaningful labels and one with abstracted labels). Our results rule out this possibility, strengthening our argument that belief is the key factor driving estimation bias.
\newline
\vspace{-2mm}

% Contributions
\noindent \textbf{Contribution:} These findings shed light on the difficulty that data communicators face when persuading people with data: as definitive and objective as the data is believed to be, our interpretation is often biased by our expectations and beliefs. 
These findings also have implications for how researchers and data analysts should approach the data analysis process. When approaching an analysis with existing hypotheses, an analyst may put themselves at risk of seeing what they want to see in their data. 
This relatively low-level bias may become compounded by higher-level ones, such as confirmation bias, which might lead to overly confirmatory follow-up studies or analyses (\cite{mynatt1977confirmation}). 

%%%%%%%%%%%%%%%%%%%%%%
\section{Related Work}
\label{section:rw}

Although the ideal visual analytic workflow is bias-free, in reality, people can fall victim to a variety of cognitive and perceptual biases when making sense of data \cite{valdez2017framework, wall2017warning, dimara2018task, xiong2019illusion, xiong2022investigating}.
Visualization readers can latch on to salient features when interpreting visualizations and gravitate toward unique colors \cite{ajani2021declutter}, larger fonts \cite{wolfe1994guided, hegarty2010thinking}, and patterns that are aligned with their beliefs and agendas \cite{xiong2019curse, NYT_Example}, giving in to letting their intuition drive decision making \cite{padilla2018decision, kahneman2011thinking}.

% Motivated Perception
\subsection{Motivated Perception}
Predictive models of perception suggest that what we see is determined by the visual input and prior knowledge \cite{geisler2002illusions} and that existing knowledge and beliefs can bias judgments even when there is less ambiguity. 
For example, in a study where people were asked to report the size of real-world objects in an otherwise dark field of view, their past knowledge of the typical size of those objects affects their perceived size and inferred distance from those objects \cite{gogel1969effect}. 
Our knowledge of real-world object size can similarly affect the perceived speed of that object in motion \cite{martin2015effect}. 
In another example, afterimages that conform to expectations are perceived as more vivid, while those that contradict expectations are less vivid \cite{lupyan2015object}.

More recently, data visualization researchers have begun exploring the extent to which existing beliefs or background knowledge can impact how people view data.
A visualization contains many patterns to perceive \cite{shah2011, xiong2019curse, xiong2018perceiving}, and viewers can perceive the patterns that are more aligned with their existing knowledge to be more salient.
For example, when participants were primed with backstories before seeing a visualization, the features mentioned in the backstories were perceived as more visually salient \cite{xiong2019curse}. 
People can also be biased by their beliefs when looking at visualizations \cite{luo2021attentional, luo2019motivated}. 
Those who do not believe in climate change tend to look more at the flat sections, confirming their thoughts on climate change.

While these effects corroborate findings in basic human perception, it is not clear whether objective, low-level visual properties can also be biased by our beliefs. 
Low-level visual properties, such as brightness, follow Weber's law, where the just-noticeable difference is a fixed proportion of its magnitude \cite{coren2004sensation}.
Among these properties, correlation perception is a key property relevant to visual analytics \cite{rensink2017, amar2005low}. Thus, as an initial step, we empirically examined how prior beliefs may influence correlation estimations.

%%%%%%%%%%%%%%%%%%%%%%%%%%%%%%% Belief Capturing 
\subsection{Beliefs and Updating}

When we interpret visualized data, our internal representations of relevant knowledge play a critical role \cite{padilla2018decision, liu2010, trafton2005, larkin1987, hegarty2004}.
When people encounter new information, they may struggle to properly incorporate past knowledge and existing beliefs to make judgments with regard to the new information, leading to biases in data interpretation \cite{dieckmann2017seeing}.
For example, failed belief updating can drive people to be overconfident in their judgment \cite{fischhoff1982,tversky1971}, as can confirmation bias \cite{tversky1971, kahneman2011thinking}.
People who are subject to confirmation bias often attend and search only for information that supports prior beliefs and do not consider alternative explanations \cite{nickerson1998confirmation, wason1968reasoning, klayman1987confirmation}. 
They may also over-weigh information congruent with their beliefs while under-weighing information incongruent with their beliefs \cite{klayman1995varieties}. 

% Showing uncertainty information can help people update properly
Visualization researchers have identified visualization techniques to facilitate adequate belief-updating \cite{mahajan2022vibe}.
For example, effectively showing uncertainty in data, such as using hypothetical outcome plots \cite{hullman2015hypothetical, kale2018hypothetical}, gradient plots \cite{correll2014error}, and quantile dot plots \cite{kay2016ish, kale2020visual, castro2021examining}, has been found to help people properly update their beliefs \cite{kim2020designing}. 
However, most visualization practitioners hesitate to incorporate uncertainty information in their designs due to concerns about their ability to precisely represent the uncertainty in a way readers can understand \cite{hullman2019authors}.
This motivates us to investigate and model the effect of beliefs with more commonly used scatterplots.%, identify alternative solutions that might help people update beliefs and mitigate biases introduced by their existing beliefs.

% But most visualziations in the wild don't show uncertain inforamtion, and when tehy do, it's dificult for people to properly make sense of it.
% thus it's imporatnt to model how people might be impacted by their beleif when making sense of sipmle visualizations.

% Thus, inspired from the study that found confirmation bias for data shown in table format \cite{Kahan2017Motivated}, we investigate whether a similar task would reveal confirmation bias in different visualizations of those data.

%%%%%%%%%%%%%%%%%%%%%%%%%%%%%%%%%%%%%%%%%%%%%%%%%%
\subsection{Scatterplots and Correlation Perception}

Current taxonomies in visualizations systematically abstract visual analytic tasks into granular operations (e.g.,\cite{brehmer2013multi, gotz2008empirical}). 
To our knowledge, the lowest levels to date include: Retrieve Value, Filter, Compute Derived Value, Find Extremum, Sort, Determine Range, Characterize Distribution, Find Anomalies, Cluster, Correlate \cite{amar2005low}.

Finding correlation is a commonly used low-level task in visual analytics \cite{liu2021data} that has been well-studied in the visualization community \cite{rensink2010perception, harrison2014ranking, kay2015beyond}.
Scatterplots are shown to especially facilitate correlation extraction in visualization \cite{saket2018task}.
Human participants are sensitive to correlation values and can generally read correlations from scatterplots accurately with little training \cite{rensink2010perception}.

Correlation perception is, of course, imperfect and is likely perceived using cues other than correlation per se \cite{rensink2017}. 
Recent work argues such perceptual shortcuts might include using the aspect ratio of the probability density function generated by the dots as a proxy for correlation (usually in the form of an ellipse that describes a constant probability of encountering a dot) \cite{harrison2014ranking, yang2018correlation, kay2015beyond}. 

Changes to the design of the visualization can also impact how people extract key statistics \cite{ gleicher2013perception}.
For example, the shape and orientation of the visual encoding marks can impact correlation perception in scatterplots, such that participants are more accurate if the direction of the shapes aligns with the regression line \cite{liu2021data}.  
The size and lightness of the scattered dots can distort visual summaries of data and lead to misjudgments in their weighted average \cite{hong2021weighted}.
These findings suggest that although accurate, there is room for bias to nudge correlation estimations when people read scatterplots. 
Together with the prevalence of scatterplots and correlation tasks in visual analytics, they make the correlation estimating in scatterplots an ideal test bed to model the potentially biasing effect of belief. 

% Indeed, recent work has shown that
% judgements of correlation vary widely in accuracy and speed depending
% on a visualization’s format, consistent with the idea that the visual
% system does not extract correlation per se, but instead relies on some set of perceptual tasks that serve as proxies for correlation (\cite{harrison2014ranking, kay2015beyond, rensink2010perception}).

% But our perception can be biased by other factors like average dot size \cite{hong2021weighted}.
% Mean judgments can also be impacted by saliency of visual encoding
% Judgements are no harder when each set contains more points, redundant and conflicting encodings, as well as additional sets, do not strongly affect performance, and judgements are harder when using less salient encodings. \cite{}

% visual encoding mark orientation can impact correlation perception. More accurate if aligned with regression line direction. 

%%%%%%%%%%%%%%%%%%%%%%
\section{General Method}
The stimulus, data, and analysis script are available at the Open Science Framework website: \url{https://osf.io/3cnza/}. %\url{https://osf.io/3cnza/?view_only=4db249b803ce4cfdb713a3bfd2dfeb79}.

\subsection{Stimuli}
\begin{figure}[t!]
\centering
 \includegraphics[width = \columnwidth]{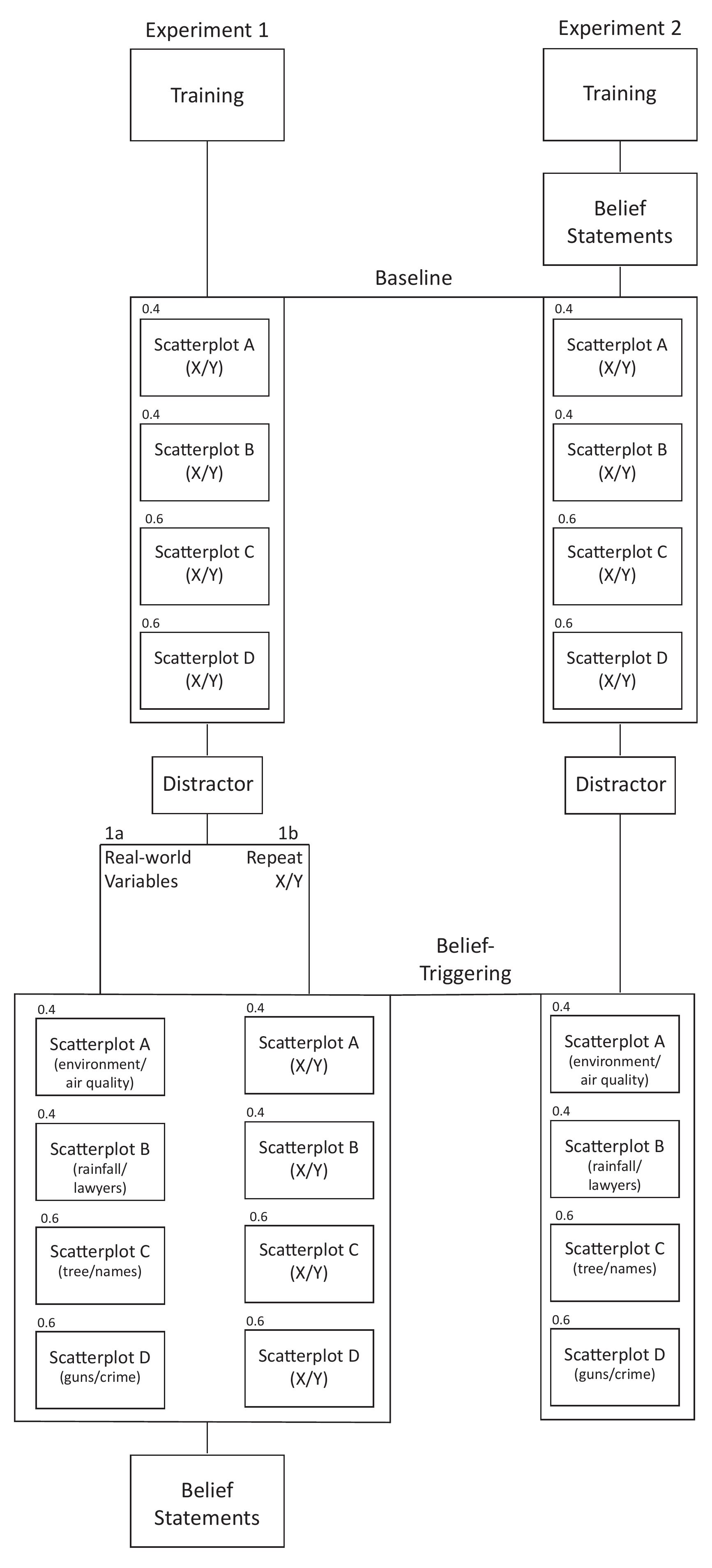}
 \caption{Flowchart depicting the procedure for each experiment. %Experiment 1a surveyed participants' beliefs at the end of the experiment, while Experiment 2 surveyed participants' beliefs in the beginning of the experiment. In Experiment 1b, instead of having alphabetic labels `X' and `Y' on the axes, different letters were used to label each scatterplot to differentiate the plots shown before and after the distractor session
 }
 \label{fig:Exp_Method_Flow}
\end{figure}

\begin{figure}[t!]
\centering
 \includegraphics[width = 0.9\columnwidth]{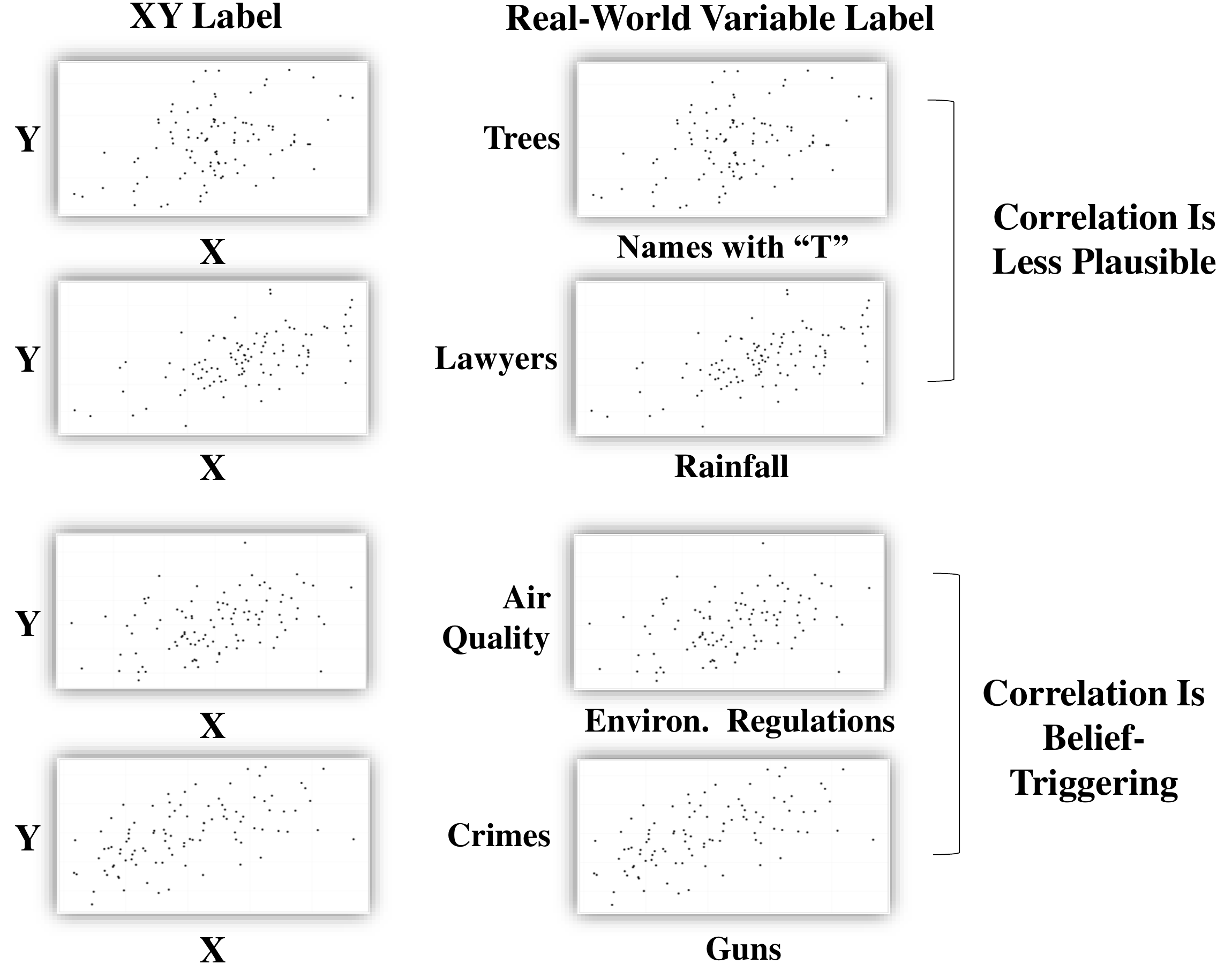}
 \caption{Scatterplots used in our experiments, either labeled with letters `X' and `Y' or with real-world variable names, depicting either a 0.4 or a 0.6 correlation. Note the two scatterplots in the same row contains identical data. %Two sets of the real-world variables are belief-triggering, and the other two less plausible. Participants viewed the scatterplots twice, before and after completing a distractor task. 
 The aspect ratio of the scatterplots was approximately 2:1 (1399x712 pixels) and the size of the scatterplots was adjusted to 50\% of the width of the participant's screen during the survey. Axes values were not included, as to eliminate the possibility of calculation (or estimated calculation) of correlation.}
 \label{fig:SameScatterplotsDifferentLabels}
\end{figure}

Stimuli for these experiments consisted of four scatterplots, with values randomly generated to fit a given correlation. In order to provide participants with ample space to over- or under-estimate the correlation, we chose intermediate correlation values of 0.6 and 0.4. Among the four, two scatterplots had a correlation of 0.4, and the other two had a correlation of 0.6. The axes titles on each scatterplot varied within experiments, but the datasets were identical across control and experimental stimuli (see Figure \ref{fig:SameScatterplotsDifferentLabels}). The survey was hosted by Qualtrics \cite{qualtrics2013qualtrics}, and the survey flow for each experiment can be seen in Figure \ref{fig:Exp_Method_Flow}. Neither exact stimuli nor the correlation values were shown in the training portion of the surveys. 

\subsection{Participants}
Among all experiments, 748 adults were recruited from an online crowdsourcing site, Amazon’s Mechanical Turk (MTurk) \cite{heer2010}. Of these 748, 435 were men, 306 were women, 5 were non-binary, and 2 chose not to disclose their gender. The average age of all participants was 40 years old (SD = 10.62 years). To be included in the analysis, responses had to have the correct answer to all attention check questions and participants must not have seen any of the questions besides demographics in any previous survey. Specific attention check questions are described in Section \ref{section:procedure}.

This led to 27 exclusions in Experiment 1a, 200 exclusions in Experiment 1b, and 32 exclusions in Experiment 2. We recruited `Master' workers for Experiment 1a and Experiment 2, but due to difficulty collecting responses with `Master' respondents for Experiment 1b (responses came in at an extremely slow pace), we removed this requirement for Experiment 1b. An additional attention check was used for Experiment 1b to ensure that participants' response and attention quality remained comparable to `Master' workers. For this reason, the number of exclusions in Experiment 1b was notably high.

% about pay
Participants completed a Qualtrics survey %detailed in the following section 
and were paid between \$1.35 and \$2.00, depending on the number of sections in the survey, at a rate of \$9.47 an hour. This research adhered to ethical guidelines including the legal requirements of the United States as well as the Institutional Review Board. 
% Further details about participants can be found in each experiment's respective section. 

\subsection{Procedure}
\label{section:procedure}

As shown in Figure \ref{fig:Exp_Method_Flow}, in all experiments, participants completed a similarly structured survey. 

The first attention check question occurred prior to the training block. The page read ``Thank you for participating in our survey. We will ask you to answer short questions about \textbf{graphs.} Please read \textbf{all} questions and answers carefully before responding.'' Participants then moved to the next page and were asked to fill in the sentence ``We will ask you to answer short questions about \_\_\_.''

Following this check, all participants were trained to extract correlation from scatterplots through an explanation of correlation, followed by seven examples, ranging from -0.9 to 0.9, labeled with the correct correlation value. They then completed three practice estimations during which they received feedback about any error in estimation.

The second attention check came just before viewing the first experimental stimuli. Participants were informed on the previous page, ``Please estimate the correlation of the following graphs. This is different from the previous section, as you will \textbf{not} receive feedback upon selecting your answer.'' They were then asked, ``Will you receive feedback after selecting your answer?'' and given options of `yes' and `no'.

After this attention check, participants estimated the correlation of four scatterplots, all labeled with `X' and `Y', using a sliding scale from 0 to 1. Two of the scatterplots contained different datasets, both depicting a correlation of 0.4, and the other two depicted a correlation of 0.6.
The same four scatterplots were viewed by all participants in a randomized order of display.  

After completing the correlation estimations on the first set of four scatterplots, participants completed a distractor task. 
%The distractor task consisted of reading simple bar charts and extracting conclusions in sentence form. 
In Experiment 1a and Experiment 2, participants then went on to complete estimations of the same four scatterplots as the previous estimations, but with real-world variables instead of `X' and `Y' on the axes, as shown in Figure \ref{fig:SameScatterplotsDifferentLabels}.
The participants always viewed the `X' and `Y' labeled scatterplots first, to ensure that they were not primed by real-world variables.
This allowed us to treat their estimation in the `X' and `Y' labeled scatterplots as an unbiased baseline.

Two of these pairs were topics that were more likely to trigger beliefs (gun ownership and violent crimes; environmental regulations and air quality), and two were neutral topics not usually associated with strong beliefs (the number of trees and people whose names start with `T;' the number of lawyers and annual rainfall). 
Including both belief-triggering and neutral topics provided us the ability to examine the effect of belief not only on topics that are belief-triggering but also on those that seem less plausible.

We also counterbalanced the correlation values with the real-world variable labels through two conditions to ensure the chart topic and its correlation values were not confounded. 
In condition 1, the plot depicting gun ownership and the plot depicting the number of trees had correlations of 0.6. The other two topics were depicted with correlations of 0.4.
In condition 2, the correlation values were swapped between the topic pairs. % belief-trigger and neutral variable pairs.
%plot previously depicting gun ownership had the environmental regulation labels, and vice versa. The plot previously depicting the number of trees had the annual rainfall labels, and vice versa. 

In Experiment 1a and Experiment 2, participants also rated their level of agreement with statements representing a belief in relationships between the variables depicted in the scatterplots, noted as ``Belief Statements'' in Figure \ref{fig:Exp_Method_Flow}. 
For example, participants rated the statement, ``I believe that as the number of households with guns increases, the number of violent crimes increases,'' on a 7-point Likert scale.
A rating of 1 indicated that the participant strongly disagreed with the statement, while a rating of 7 indicated that the participant strongly agreed with the statement. 
Similar Likert scales have been used to measure attitudes and beliefs \cite{joshi2015likert, hale2012measuring, wilkins2008relationship}. 
We were specifically interested in their belief in a correlation between the two variables shown and thus used the same phrasing as the scatterplot axes labels. The plots all presented positive correlations to maximize the comparability of results across examples.

Participants in Experiment 1a made these ratings after completing a second distractor task following the correlation ratings of all scatterplots, not shown in Figure \ref{fig:Exp_Method_Flow}, to further the distance between seeing a visual depiction of the scatterplots and reporting their belief on the topic. 
All participants in Experiment 2 made these ratings in the beginning, after completing the correlation training section but before estimating the correlations for any scatterplots. 
A subset of participants in Experiment 2 also completed the same ratings at the end of the survey, as a measure of personal belief for each participant.

In Experiment 1b, participants completed the distractor task after rating the correlation of the first four scatterplots labeled with `X' and `Y'. 
After the distractor task, instead of seeing the same scatterplots with real-world labels, they saw them with a different set of abstracted letter labels, such as `M' and `N'.
Participants in this experiment did not complete any ratings of belief. There was only one condition of this experiment, with each set of scatterplots shown in a randomized order.

Overall, Experiment 1a and Experiment 2 examined the effect of belief on correlation estimations. 
% Experiments 1a and 2 always showed the scatterplots with real-world variables after those with alphabetic labels.
Experiment 1b served as a control condition for Experiment 1a, to account for the possibility that differences in correlation ratings between the two sets of scatterplots were a result of time delay.

%%%%%%%%%%%%%%%%%%%%%%%
\section{Experiment 1a Measuring Belief Last (Posterior)}

In this experiment, as shown in Figure \ref{fig:Exp_Method_Flow}, participants completed training on estimating correlation, and then estimated the correlation for four scatterplots with `X' and `Y' labels on the axes (correlations of 0.4 and 0.6), as shown in Figure \ref{fig:SameScatterplotsDifferentLabels}.
Next, they completed a distractor task before rating the same four scatterplots, this time with real-world variables on the axes. 
After a second distractor task, participants rated their level of agreement with statements depicting belief in the relationship between variables in the scatterplots. 
In the end, participants reported demographic information such as their age and level of education. 

\subsection{Participants}
% Power Analysis
Assuming a medium effect size (a $\rho$ = 0.21 correlation based on pilot studies) for the effects of belief, power analysis with G*Power \cite{faul2007g} suggests that a target sample of 291 participants would give us 95\% power to detect an overall effect of the belief factor at an alpha level of 0.05. 
After applying the exclusion criteria specified previously, we ended up with 295 participants (172 men, 120 women, and three non-binary people) with an average age of 41 years (SD = 9.66 years).

\subsection{Ability to Estimate Correlations} 
We first examine how accurately participants estimated the correlation values in the scatterplots without considering the effects of belief. Overall, participants were quite accurate at estimating correlations in scatterplots, corroborating existing findings. 

We compared participant estimations of the XY labeled scatterplot correlations with the ground truth correlations (which is either 0.4 or 0.6). Participants' correlation estimates for the two versions of the 0.4 and 0.6 correlation scatterplots slightly differ from ground truth, but generally, they were quite accurate. % in reporting the correlation values in these XY labeled scatterplots. 
They reported an average correlation of 0.415 (SD = $0.19$, CI = [$0.394$, $0.437$]) and 0.401 (SD = $0.21$, CI = [$0.377$, $0.425$]) for the two versions of scatterplots with a 0.4 correlation. For scatterplots with a 0.6 correlation, participants reported an average correlation of 0.558 (SD = $0.19$, CI = [$0.537$, $0.579$]) and 0.625 (SD = $0.18$, CI = [$0.604$, $0.646$]). 

% How well did people do for real-world labeled?
We compared the correlation estimates of the same scatterplots with and without real-world labels to identify potential bias resulting from labeling the axes with real-world variable pairs. 
Participants reported significantly different correlation values for the same scatterplot in the presence of real-world variable pair labels. 
On average, they estimated the correlation for the scatterplots with a 0.4 correlation with real-world variable labels to be 0.385 (SD = $0.19$, CI = [$0.363$, $0.407$]) and 0.287 (SD = $0.22$, CI = [$0.262$, $0.312$]). They estimated the correlation for the scatterplots with a 0.6 correlation with real-world labels to be 0.561 (SD = $0.19$, CI = [$0.539$, $0.583$]) and 0.546 (SD = $0.23$, CI = [$0.520$, $0.572$]). 

%%%%%%%%%%%%%%%%%%%%%%%%%%%%%%%%%%%%%%%%%%%%%%%%%%%%%%%%%%%%%%%
% \subsection{Not Plausible vs Belief-Triggering Plots}
% To test whether participants made differing estimations for the same scatterplot when it was labeled with ``X'' and ``Y'' compared to when it was labeled with real-world variables, we computed the differences between the two estimations. 
% A two-sided paired-sample t-test revealed that there is a significant difference between participants' correlation estimations when they viewed scatterplots with XY labels and when they viewed plots with real-world variable pairs (MD = $-0.055$, t = $-8.38$, p < $0.001$). 

% In the following sections, we examine how this difference may be associated with how much they believe or agree that the real-world variables are correlated. 

% what are people distribution in belief like?
\begin{figure*}[th]
\includegraphics[width = \textwidth]{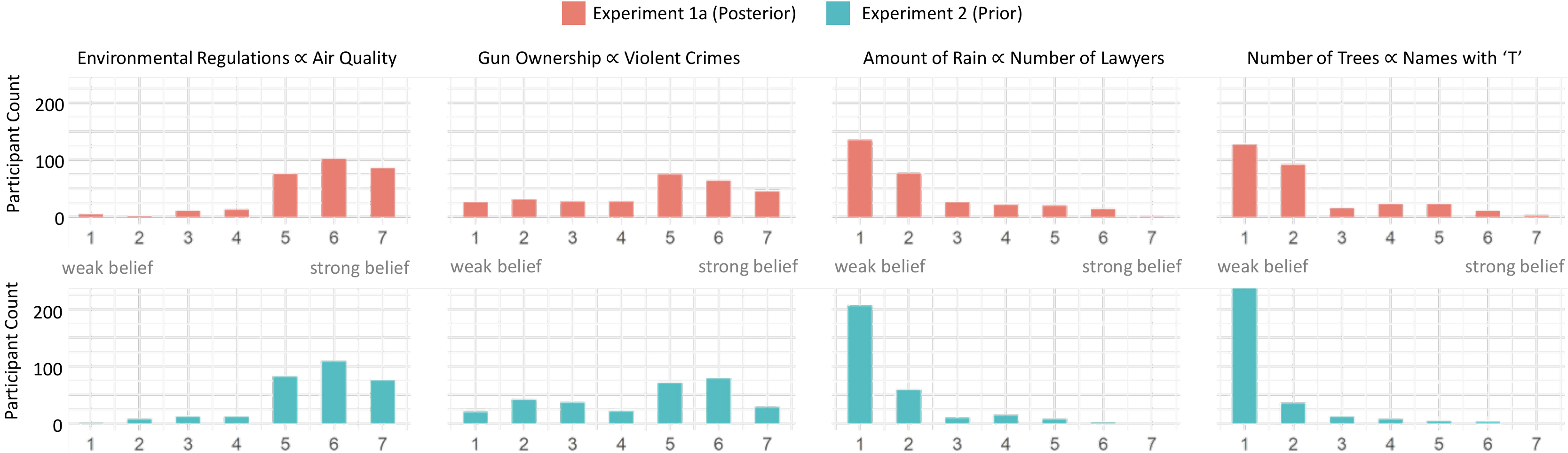}
 \caption{Distribution for the level of participant agreement with statements supporting a positive correlation between the variables depicted in the scatterplots. This was measured at the end of Experiment 1 (Top) and at the beginning of Experiment 2 (Bottom). The gun/crime and environment/air topics have right-skewed distributions such that more people strongly agreed with the positive correlation statement, while the trees/names and lawyers/rainfall topics have left-skewed distributions such that more people strongly disagreed with the positive correlation statement. }
 \label{fig:Exp1a2_BeliefDistribution}
\end{figure*}

% How does belief correlate with estimation error?
\begin{figure*}[h]
\includegraphics[width = \textwidth]{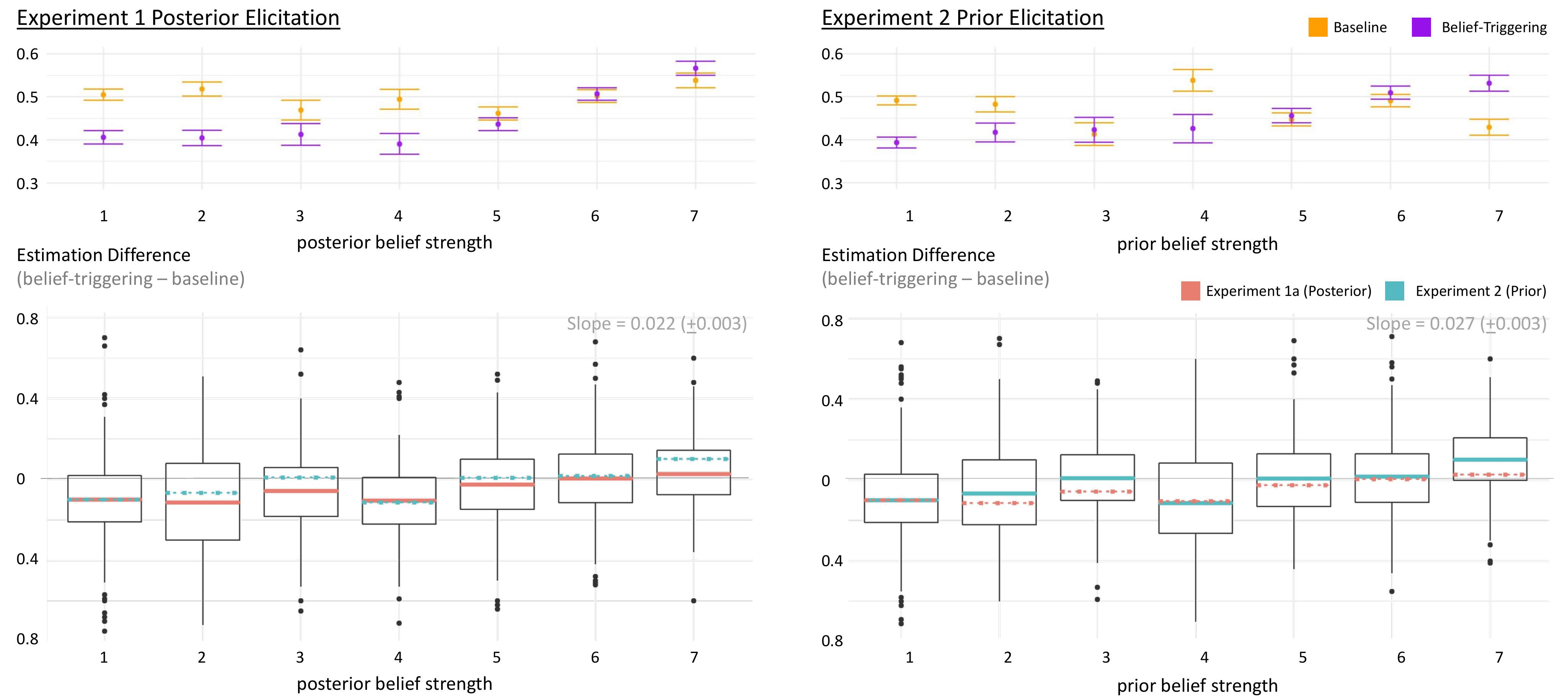}
 \caption{Top: Shows the raw estimations for scatterplots labeled with baseline (XY) and belief-triggering real-world variables. Bottom: Shows how the difference in correlation estimations between baseline (XY) and belief-triggering plots varies with the strength of belief. %Each dot represents the difference in correlation estimation between the same scatterplot, labeled with XY or real-world variables. Each participant contributes 4 data points on the chart. Greater than 0 on the y-axis means that participants overestimated the correlation. 
 Less than 0 means that participants underestimated the correlation. The colored horizontal lines represent the average values per belief strength. Both posterior and prior averages are shown in both charts to facilitate comparisons between them. Left: Experiment 1 where the strength of belief is elicited at the end of the experiment. Right: Experiment 2 where the strength of belief is elicited towards the beginning of the experiment. For both experiments, participants with stronger beliefs more likely overestimated the correlation for the real-world variable labeled scatterplots.}
 \label{fig:RegressionsExp1Exp2}
\end{figure*}

% Beliefs for plausible and belief triggering topics
\begin{figure}[t!]
\centering
 \includegraphics[width = \columnwidth]{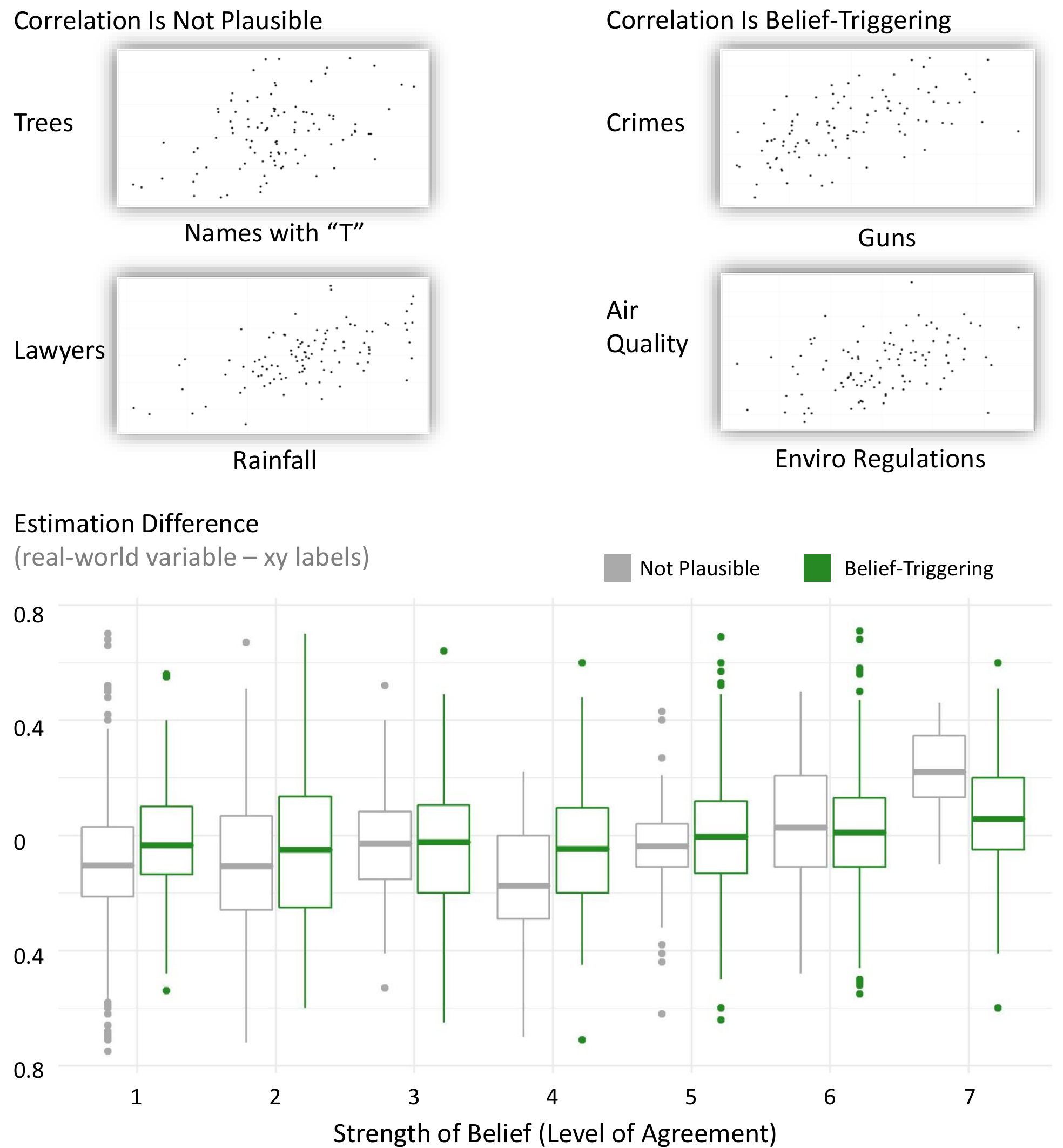}
 \caption{This belief-driven over and underestimation of correlations happens when people see both not plausible real-world variable pairs (e.g., lawyers and rainfall) and belief-triggering variable pairs (e.g., guns and crimes). Data is consolidated from both Experiments 1a and 2.}
 \label{fig:BeliefTriggering_NotPlausibleComparison}
\end{figure}

%%%%%%%%%%%%%%%%%%%%%%%%%%%%%%%%%%
\subsection{Self-Reported Beliefs}
To examine the effect of participant belief on their correlation estimates, we showed them two plots with belief-triggering variable pairs (gun ownership and violent crimes; environmental regulations and air quality), and two with neutral variable pairs not usually associated with strong beliefs (the number of trees and people whose names start with ``T;'' the number of lawyers and annual rainfall). 

This analysis verifies that participants more strongly believed that there exists a positive correlation between gun ownership and violent crimes (M = $5.72$, SE = $0.07$) and between environmental regulations and air quality (M = $4.57$, SE = $0.11$). 
They less strongly believed that there exists a positive correlation between the number of trees and people whose names start with ``T'' (M = $2.19$, SE = $0.088$) and between the number of lawyers and annual rainfall (M = $2.21$, SE = $0.089$). 
Post-hoc pair-wise comparisons between their belief on the four topics suggest that participants most strongly believed in the gun relationship, followed by the environment relationship, followed by the trees/names, and the lawyers/rainfall relationship (their beliefs of which are not statistically different, p = $0.998$). 
Figure \ref{fig:Exp1a2_BeliefDistribution} shows the belief distributions for the four topics.

% Participants' distribution for the level of agreement with statements depicting a positive correlation between the variables depicted in the scatterplots at the end of the experiment. The gun/crime and environment/air topic have right skewed distribution such that more people strongly agreed with the positive correlation statement, while the trees/names and lawyers/rainfall topic have left skewed distribution such that more people strongly disagreed with the positive correlation statement.

% Effect of belief on correlation estimate
\subsection{Effect of Belief on Correlation Estimates}
To investigate whether labeling the scatterplot with real-world variables biases participants' correlation estimations, we compare the differences between their estimation for the same scatterplot labeled with `X' and `Y' and with real-world variables. 
We demonstrate this difference to be correlated with how strongly they believe the real-world variables are positively correlated. 

We constructed a mixed-effect linear model predicting the difference between estimates in the scatterplots labeled with `X' and `Y' and with real-world variables using the lmer function in R \cite{bates2007lme4}.
We used experiment versions (counterbalanced conditions), topics (the two belief-triggering and two neutral topics), the strength of belief (the degree of agreement with the correlation statements), and the interaction between topics and the strength of belief as predictors, and included a random effect of participants. 
We also added demographic variables including age, gender, political orientation, level of education, experience with statistics, and experience with doing research as fixed effects to explore potential differences.

We found a significant main effect of belief (${\chi}^2$ = $18.39$, Est = $0.024$, SE = $0.011$, t = $2.23$, p = $0.023$), such that the stronger the participant believes in the variables' correlation, the more they overestimated the correlation when the scatterplot was labeled with that variable pair, as shown in Figure \ref{fig:RegressionsExp1Exp2}. 
Similarly, the more they disagreed with the variable correlation, the more they underestimated the correlation. 

There was no significant main effect of condition (${\chi}^2$ = $0.022$, p = $0.88$), suggesting that the two counterbalanced conditions were not different from one another.
There was also no significant main effect of topic (${\chi}^2$ = $4.28$, p = $0.23$), nor a significant interaction between topic and belief (${\chi}^2$ = $1.84$, p = $0.61$). 
This suggests that regardless of whether the topic was belief-triggering or neutral, the under/overestimation of the correlation was driven by the participants’ own beliefs. 
Figure \ref{fig:BeliefTriggering_NotPlausibleComparison} shows a comparison of the relationship of correlation estimation error and belief strength, between belief-triggering topics and less plausible topics. The two look extremely similar.

We also found no significant effect of any demographic variables. Details can be found in the supplementary materials.

%%%%%%%%%%%%%%%%%%%%%%%%%%%%%%%%%%%%%%%%%%%%%%%%
\subsection{Ordering Effects and Alternative Explanations}
Despite accurately estimating scatterplot correlations when the plots were labeled with `X' and `Y', participants became less accurate when estimating the correlation for the same scatterplots but with real-world variable labels. 
Our model suggests that the deviation in correlation estimation is associated with the strength of participant beliefs, such that strong beliefs predict overestimation and strong disbeliefs predict underestimation. 

However, one possible alternative explanation for the over/underestimation could be the factor of time. 
Participants always saw the real-world labeled scatterplots after they had seen the `X' and `Y' labeled scatterplots, at which point they could have become habituated or anchored by previously seen scatterplots.
Further, more time has lapsed between training and estimating correlations for the real-world variable labeled scatterplots, compared to the time between training and estimating correlation for the `X' and `Y' labeled scatterplots.
So it is possible that the effect of training has deteriorated, and thus participants provided less accurate correlation estimations. 
This motivated us to conduct Experiment 1b, a control experiment where, instead of showing participants the `X' and `Y' labeled scatterplots followed by the same plots labeled with real-world variables, we showed them the same `X' and `Y' labeled scatterplots twice to account for the potential effect of time.

%%%%%%%%%%%%%%%%%%%%%%%%%%%%%
\section{Experiment 1b Controlling for the Effect of Time}
We observed a significant difference in the correlation values participants reported for the scatterplots labeled with `X' and `Y', compared to the same scatterplots labeled with real-world variables. 
We investigate whether the difference is driven by the real-world labels, or simply by the passage of time in this experiment.

\subsection{Method}
Participants completed training on rating correlation, then rated four scatterplots with alphabetic labels (e.g., X, Y), with correlation values of either 0.4 or 0.6. 
They then completed a distractor task before rating the same four scatterplots, also with alphabetic labels on each axis. 
The alphabetic labels are used to strip context away from the scatterplots to ensure that belief does not play a role in our results.
Participants then reported demographic information. 

\subsection{Participants}
Because Experiment 1b is intended as a control version of Experiment 1a with no manipulation before and after the distractor task, we recruited participants to match the number of participants in one condition of Experiment 1a, which is equivalent to half the number of participants from Experiment 1a. After applying the exclusion criteria specified previously, we ended up with 153 participants (86 men, 64 women, 1 non-binary person, and 2 who chose not to disclose), with an average age of 38 years (SD = $11.49$ years).

\subsection{Results}
% \subsubsection{Summary Statistics}
%  on Control Phase 1 and 2 + comparing them
Participants estimated the correlation of scatterplot with alphabetic labels twice (e.g., `X' and `Y') in Experiment 1b, which we will refer to as Phase 1 and Phase 2 estimates. 
In Phase 1, the average correlation they reported for the two versions of the 0.4 correlation scatterplots were 0.423 (CI = [$0.395$, $0.451$]) and 0.354 (CI = [$0.395$, $0.451$]). The average correlation they reported for the two versions of the 0.6 correlation scatterplots were 0.541 (CI = [$0.509$, $0.572$]) and 0.599 (CI = [$0.570$, $0.628$]).
In Phase 2 the average correlation they reported for the two versions of the 0.4 correlation scatterplots are 0.442 (CI = [$0.413$, $0.472$]) and 0.372 (CI = [$0.339$, $0.406$]). The average correlation they reported for the two versions of the 0.6 correlation scatterplots are 0.513 (CI = [$0.482$, $0.545$]) and 0.629 (CI = [$0.600$, $0.656$]).

We compared participants' correlation estimates for the real-world variable labeled scatterplots in Experiment 1a to the Phase 2 estimations in Experiment 1b (labeled with alphabetical letters). 
Between these two sets of estimates, the only difference is how the scatterplots were labeled. 
If the two sets are similar, that suggests that timing is likely the factor driving any deviation from their estimation of the first set of scatterplots labeled with `X' and `Y' (phase 1). 
If the two are different, it likely suggests that the real-world variable labels, which trigger either belief or disbelief in a positive correlation, affected correlation estimations, rather than it being an effect of time. 

We found there to be a significant difference between them. 
Pair-wise comparisons with Bonferroni corrections suggest that the correlation estimates for the two versions of 0.4 correlation plots were significantly different ($p_{v1}$ = $0.002$, $p_{v2}$ < $0.001$), and that of the two versions of the 0.6 correlation plots were also significantly different ($p_{v1}$ = $0.016$, $p_{v2}$ < $0.001$). 
We can infer from these comparisons that the deviations in correlation estimates with the scatterplots labeled with real-world variables in Experiment 1a are not caused by time delays, but rather by the real-world labels and their associated beliefs.

% Transition to Experiment 2 - Prior Belief rather than Posterior
In both Experiment 1a, participants viewed the real-world variable labeled scatterplots before reporting their belief (level of agreement with a positive correlation between the depicted variables). 
It is possible that these beliefs that people reported are posterior in nature such that they were not their true beliefs, but rather updated beliefs affected by the scatterplots they saw. 
To demonstrate that participants' prior beliefs have an influence on how they estimate correlations in scatterplots, we conduct Experiment 2 where we elicited their level of agreement on the depicted topics prior to them seeing the real-world variable labeled correlations.

%%%%%%%%%%%%%%%%%%%%%%%%%%%%%%%%%%%
\section{Experiment 2 Measuring Beliefs First (Prior)}
In Experiment 1a, we measured participants' beliefs on relevant real-world variable pairs at the \textit{end} of the experiment, after they had seen scatterplots depicting potential relationships between those variables. 
To account for the potential effect of these visualizations on participants' beliefs, we conducted Experiment 2 and measured their beliefs on these real-world variable pairs at \textit{the very beginning} of the experiment. 

\subsection{Method}
Similar to both previous experiments, participants first completed training on correlation estimation. 
Immediately following training, they rated their level of agreement with belief statements regarding the variables previously viewed in the scatterplots. 
They then rated the correlation of four scatterplots with `X' and `Y' labels on the axes (correlations of 0.4 and 0.6). 
Next, they completed a distractor task before rating the same four scatterplots, this time with real-world variables on the axes. 
Finally, participants reported demographic information such as their age and level of education. 

% Power Analysis
\subsection{Participants}
Assuming a medium effect size (a $\rho$ = 0.21 correlation based on pilot studies) for the effects of belief, power analysis with G*Power \cite{faul2007g} suggested a target sample of 291 participants for 95\% power to detect an overall effect of the belief factor at an alpha level of 0.05. After applying the exclusion criteria specified previously, we ended up with 300 participants, with an average age of 40 years (SD = $11.01$ years). 

\subsection{Ability to Estimate Correlations}
% \yeaseul{add pair-wise t.test}
% how well did people do for original (XY)?
% CX: turns out the two versions for 0.6 were significantly different so looking at the four plots separately 
To gauge how accurate participants were in estimating of `X' and `Y' labeled scatterplot correlation, we compared their response with the ground truth correlations. As in Experiment 1a, the correct answers were either 0.4 or 0.6. 
We observed that participants overall performed well on estimating the correlation. On average, participants reported 0.381 (SD = $0.18$, CI = [$0.36$, $0.40$]) for the first 0.4 stimuli, and 0.38  (SD = $0.21$, CI = [$0.35$, $0.40$]) for the second. For the stimuli depicting 0.6, participants reported 0.530 (SD = $0.19$, CI=[$0.51$, $0.55$]) and 0.62 (SD = $0.18$, CI = [$0.60$, $0.64$]), respectively. This was similar to the results of Experiment 1a. %, with a gap between average responses. 

Next, we compared the participants' estimations from when they saw the plots with and without real-world labels. We found that the responses were significantly different. Participants estimated the correlation for the scatterplots with a 0.4 correlation with real-world variable labels to be 0.407 (SD = $0.21$, CI = [$0.38$, $0.43$])  and 0.255 (SD = $0.22$, CI = [$0.23$, $0.28$]). They reported the correlation for the scatterplots with a 0.6 correlation with real-world labels to be 0.546 (SD = $0.20$, CI = [$0.52$, $0.57$]) and 0.546 (SD = $0.25$, CI = [$0.52$, $0.57$]).

% %%%%%%%%%%%%%%%
% \subsection{Difference between baseline and belief-triggering scatter plot estimations}
% To investigate whether there is a difference in estimating the same scatterplot when it was labeled with XY compared to when it was labeled with real-world variables, we compared the differences between the two estimations. Our analysis shows that there is a significant difference between participants' correlation estimations when they viewed scatterplots with XY labels and when they viewed plots with real-world variable pairs (MD = $-0.037$, t = $-5.504$, p < $0.001$). 

%%%%%%%%%%%%%%%%%%%%%%%%%%%%%%%%%%%%%%%%%%
\subsection{Self-Reported Beliefs}
Similar to Experiment 1a, we analyzed the influence of the beliefs on estimating correlation. 
We found that participants more strongly believed in a positive correlation between environmental regulations and air quality (M = $5.66$, SE = $1.20$) and between gun ownership and violent crimes (M = $4.45$, SE = $1.80$). 
They less strongly believed a positive correlation between the number of lawyers and annual rainfall (M = $1.55$, SE = $1.03$), and between the number of trees and people whose names start with ``T'' (M = $1.38$, SE = $0.92$). 
We ran post-hoc pair-wise comparisons with corrections between their beliefs on the four topics, and the result shows that belief ratings in the four conditions were reliably different (p < $0.0001$). Figure \ref{fig:Exp1a2_BeliefDistribution} shows the belief distributions.

%%%%%%%%%%%%%%%%%%%%%%%%%%%%%%%%%%%%%%%%%%%%%%%%%%%
\subsection{Effect of Belief on Correlation Estimates}
% Effect of belief on correlation estimate
To understand the influence of the beliefs on the correlation estimation, we ran a mixed effect linear model with the difference between their estimation on the scatterplots with and without the real-world label as a dependent variable, beliefs and the topic of the chart and their interaction as fixed effects, and participants as a random effect. 
We also added demographic variables including age, gender, political orientation, level of education, experience with statistics, and experience with doing research as fixed effects to explore potential differences.
We found that the beliefs had a reliable impact on the difference (Est = $0.025$, SE = $0.011$, t = $2.28$, p = $0.023$), as shown in Figure \ref{fig:RegressionsExp1Exp2}. 
As the average participant's beliefs were 1 unit stronger in a positive relationship of the two variables, they saw 0.02 units of additional correlation compared to when they examined the chart without the real-world labels. 
There was no reliable effect of the topic nor the interaction. Additionally, consistent with the results from Experiment 1a, there is no significant effect of topic (p = $0.085$), condition (p = $0.51$), nor the interactions between topic and belief (p = $0.21$), as shown in Figure
\ref{fig:BeliefTriggering_NotPlausibleComparison}. 
We also found no significant effect of any demographic variables. Details can be found in the supplementary materials.

% Does eliciting belief in the beginning vs. the end matter?
% \begin{figure*}[th]
% \includegraphics[width = \textwidth]{Figures/BeliefElicitationBeforeAfter.pdf}
%  \caption{This belief-driven over and underestimation of correlations happens in both Exp 1 and Exp 2, when we used posterior belief and prior belief as a predictor to estimation differences.}
%  \label{fig:BeliefElicitationBeforeAfter}
% \end{figure*}

% simulated 0.4, 0.5, 0.6 correlation for non-believers, truth, believers
\begin{figure*}[th]
\centering
\includegraphics[width = 16cm]{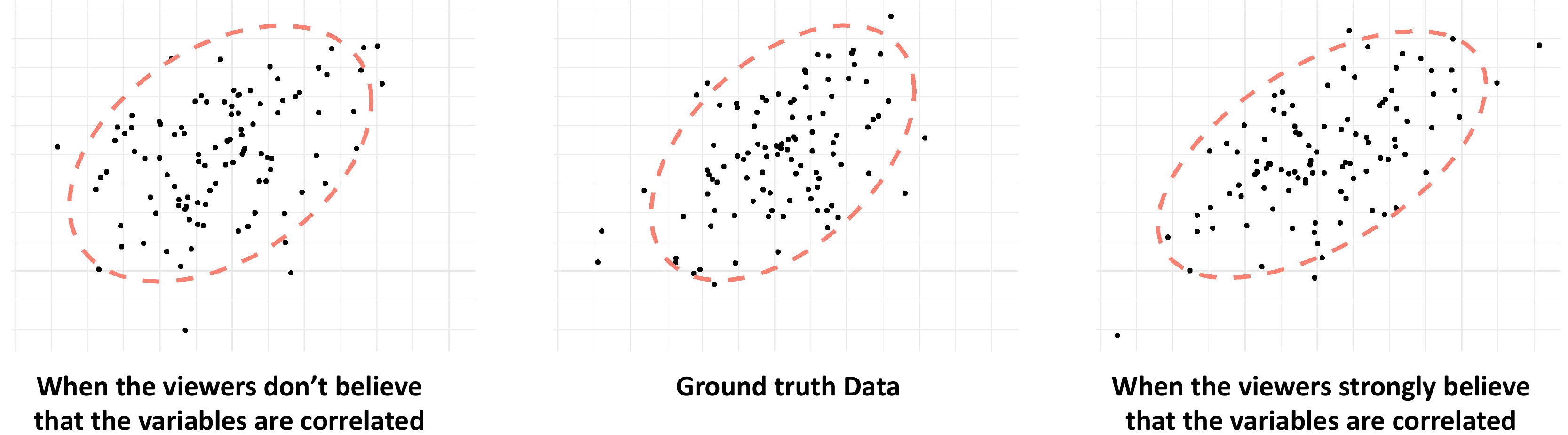}
 \caption{A visualization of the size of the bias based on effect sizes from the experiment. Middle: what the `objective' data should look like (with Pearson's correlation = 0.5). Left: How the data would look like to a viewer who doesn't believe that the variables are correlated (r = $0.4$). Right: How the data would look like to a viewer who strongly believes that the variables are correlated (r = $0.6$).}
 \label{fig:demoEffect}
\end{figure*}

%%%%%%%%%%%%%%%%%%%%%%%%%%%%%%%%%%%%%%%%%%%%%%%%%%%
\section{Comparing Experiment 1a and Experiment 2}
\label{compare12}
We compared our results from Experiment 1a and 2 to examine whether the effect of labeling scatterplots on correlation estimates replicates, and whether asking the belief questions before or after the participants see the scatterplots made a difference.
\newline
\vspace{-2mm}

\noindent \textbf{Ability to Estimate Correlations:} We did not observe a difference between participants' ability to estimate the correlation of scatterplots with `X' and `Y' labels. Since this step was shown before seeing the scatterplots with the real-world labels, we did not expect the result to differ.
\newline
\vspace{-2mm}

\noindent \textbf{Belief Distributions}
We compared the participants' beliefs when they were prompted before and after seeing the scatterplots with the real-world labels. We found no difference of their beliefs on environmental regulations and air quality (t = $-0.654$, p = $0.512$) nor gun ownership and violent crimes (t = $-0.862$, p = $0.389$). 
However, there was a difference between belief ratings for the number of trees and people whose names start with “T” (t = $-8.051$, p < $0.001$) and for the number of lawyers and annual rainfall (t = $-5.997$, p < $0.001$). 

Figure~\ref{fig:Exp1a2_BeliefDistribution} shows the belief distributions of participants in Experiments 1a and 2. 
For the belief triggering topics, there were no differences between people in Experiment 1a and 2  (Environment prior: M = $5.66$, SD = $1.201$, posterior: M = $5.722$, SD = $1.233$, Gun prior: M = $4.450$, SD = $1.800$, posterior M = $4.579$, SD = $1.864$). 
However, for the non-belief triggering topics, we observed a difference in people's belief distributions (Rain prior: M = $1.553$, SD = $1.035$, posterior: M = $2.190$, SD = $1.507$, Tree prior: M = $1.383$, SD = $0.916$, posterior: M = $2.214$, SD = $1.520$). 

\subsection{Seeing What You Believe}
% - compare effect size (of beliefs on the delta) and 
% - slope of regression line for belief (x) vs. difference between XY and real-world variable estimations (comparison of the effect of posterior and prior belief's effect size as predictors): Very similar slope, no sig differences, stats to be filled
Figure~\ref{fig:RegressionsExp1Exp2} shows the relationship between the belief strength and the correlation estimation difference. We compared the effect of prior and posterior beliefs on participants' correlation estimations when they viewed scatterplots with `X' and `Y' labels versus real-world variable pairs. The results are very similar between Experiments 1a and 2. The coefficient of the prior belief factor is 0.022, and the posterior belief factor is 0.023.

\subsection{Believing What You See}
In the case of non-belief triggering topics, we observed a slight difference in beliefs distribution when participants were asked to indicate their beliefs before and after seeing the scatterplots. This may imply that seeing the correlated scatterplots led them to believe that there might be a correlation between two variables. %Our analysis shows that the elicited beliefs can predict how much participants will deviate from the correlation estimation on XY labeled scatter plot, regardless of when the beliefs were elicited. 

%%%%%%%%%%%%%%%%%%%%%%%%%%%%%%%%%%%%%%%%%%%%%%%%%%%%%%%%%%%%%%%%%%%%%%%%%%%%%%%%%%%%%%%%%%%%%%%%%%%%%%%%%%%%%%%%%%%%%%%%%%%%%%%%%%%%%%%%%%%%%%%%%%%%%%%%%%%%%%%%%%%%%%%%%%%%%%%%%%%%%%%%%%%%%%%%%%%%%%%%%%%%%%%%%%%%%%%%%%%%%%%%%%%%%%%%%%%%%%%%%%%%%%%%%%%%%%%%%%%%%%%%%%%%%%%%%%%%%%%%%%
\section{General Discussion}
\label{disc}

% Overall Results 
People were biased by their existing beliefs when estimating correlations in scatterplots, and we were able to quantify how much those beliefs can influence the estimates. 
As shown in Figure \ref{fig:demoEffect}, people tend to underestimate the correlation of a scatterplot, by approximately 0.1 in Pearson's correlation value, when it depicts a relationship between two variables that they don't believe to be correlated. 
They tend to overestimate the correlation by approximately 0.1 when it depicts a relationship that they do believe to be correlated. 
This may explain the difficulty frequently faced when persuading people with data - as objective as the data is believed to be, our interpretations of it are often riddled with bias from our expectations and beliefs.

However, it remains unclear why participants gave differing correlation estimates for the same scatterplot when it was labeled with `X' and `Y' compared to when it was labeled with real-world variables. One explanation might be that perceived correlation strength could be impacted by `top-down' effects such that they paid more attention to scattered dots that would make the correlation look higher or lower (in accordance with their belief), discounting outliers that would strengthen or weaken the correlation against their beliefs \cite{bakker2014outlier}.
This possibility would be consistent with previous work showing that darker colored dots are weighted more when viewers estimate means in scatterplots \cite{hong2021weighted}, and from Luo \& Zhao, where viewer decisions and actions can be manipulated by making certain patterns in a visualization more visually salient \cite{luo2019motivated}. 
In this case, participants may be weighing dots that support their belief more heavily, such as weighing the dots near the regression line more heavily when they more firmly believe in the correlation.

Another possibility is that correlation estimates come with uncertainty intervals around a best guess value. When the scatterplot was labeled with `X' and `Y', viewers may be more likely to report the midpoint value in that uncertainty range. But for belief-triggering relationships, they might report a correlation either in the upper or lower range depending on the direction of their belief. 
Existing work in uncertainty demonstrates that people are more biased when the data shows more uncertainty, supporting this possibility \cite{dieckmann2017seeing, kim2020designing}. 
We discuss potential future research opportunities in Section \ref{limitation}.

%%%%%%%%%%%%%%%%%%%%%%%%%%%%%%%%%%%%%%%%%%%%%%%%%%%%%%%%%%%%%%%%%%%%%%%%
%%%%%%%%%%%%%%%%%%%%%%%%%%%%%%%%%%%%%%%%%%%%%%%%%%%%%%%%%%%%%%%%%
\section{Design Implications}
When an analyst or scientist has a belief about how the world works, their thinking can be biased in favor of that belief. Therefore, one bedrock principle of science is to minimize that bias by testing the predictions of one's belief against objective data. 
However, we have demonstrated that when people look at scatterplots, their beliefs can interact with the objective data to bias judgments of the objective strength of the relationship between two variables. 
The stronger the belief (and disbelief), the stronger the bias. 
Based on this result, when reading visualizations, if readers can first form an impression of the strength of the relationship based on only the data values (without letting the labels alone drive that impression), they might become less susceptible to belief-driven biases when interpreting the visualization. 
Similarly, data analysts should thoroughly consider their data filtering and manipulation decisions and proceed with caution once they have already seen the data. 
If the data is aligned with their beliefs, except for a couple of outliers, as shown in Figure \ref{fig:motivatingExample}, the analysts might be motivated to adjust the exclusion criteria to exclude them or to collect additional data to `drown the outliers out', which likely will skew the results and introduce biases in their analysis. 
% Outliers
% It may also demonstrate outlier exclusion bias, where you become tempted to manipulate exclusion criteria to include some data points and exclude others, and ultimately contribute to the replication crisis \cite{bakker2014outlier}.

%%%%%%%%%%%%%%%%%%%%%%%%%%%%%%%%%%%%%%%%%%%%%%%%%%%%%%%%%%%%%%%%%
\section{Limitation and Future Directions}
\label{limitation}
We discuss several limitations in our study that provide promising future research directions.
\newline
\vspace{-2mm}

% Cognition vs. Perception
\noindent \textbf{Cognition versus Perception:} We did not try to tease apart effects on perception and cognition. Making sense of data, even through a visualization, is a complex process that involves both mechanisms \cite{shah2011}. We demonstrated that belief impacts how data gets interpreted by a viewer, and quantified the extent to which it does. 

% On the perception and cognition debate
There is debate in the psychology research community on the extent to which `top-down' effects (e.g., belief or motivation) affect perception \cite{firestone2016cognition, pylyshyn1999vision}. 
Some researchers have argued that cognition has the ability to `penetrate' perception \cite{cavanagh1999cognitive, lupyan2017changing, mccauley2006susceptibility}, such that existing beliefs or personal goals affect how we perceive the natural world. 
For example, Balcetis \& Dunning demonstrated that wishful thinking could make desirable objects appear closer \cite{balcetis2010wishful}. For example, a water bottle appears closer to a thirsty person.%, money appears closer when the participant has the opportunity to win it.
As mentioned in Section \ref{section:rw}, familiarity and prior knowledge can affect size and speed perception \cite{gogel1969effect, martin2015effect}. 

However, some researchers have argued against such an interpretation of these effects, suggesting that although our knowledge and beliefs can affect what we look at or what we pay attention to, the lower-level stages of visual perception are impenetrable to existing beliefs \cite{pylyshyn1999vision, pessoa1998finding}.
%\hl{For example, Pylyshyn argued that we observe the world via an early vision system, which carries out fast but complex computations} \cite{}.
% Of course, our knowledge and expectations can still affect what we look at or what we pay attention to \cite{pylyshyn1999vision}.  %because cognitive processes influence how we allocate our attention to locations or object properties, and how we make decisions with regards to recognizing and identifying patterns after the stimulus is processed by our early vision system
% Although what we believe about the world we are looking at does depend on what we know and expect,
% we observe the world via early vision, which carries out complex computations (such as top-down processing, e.g., filling in Pessoaet al., 1998 \cite{pessoa1998finding}), 
% And this system is cognitively impenetrable.
% Cognitive penetration occurs else where - in the allocation of attention to certain locations or properties prior to the operation of early vision, and in the decisions involved in recognizing and identifying patterns after the operation of early vision.
% Seeing is not the same as believing, because 
% Visual perception is too fast to be
% impenetrable to cognition \cite{bullier1999visual}
Recent support for this idea includes a study that demonstrates the influence of object familiarity on perceived distance found in \cite{martin2015effect} % Mart{\'\i}n, Chambeaud, \& Barraza \cite{martin2015effect} 
could not be conceptually replicated \cite{mischenko2020examining}. 
Two manipulations differed between the studies: 1) instead of using an adjustment task with unlimited viewing duration, Mischenko et al. used brief presentations to minimize the potential effect of deeper thinking during task completion, and 2) instead of showing participants objects that varied significantly in size (e.g., tennis balls and golf balls), they showed participants different types of coins with similar sizes \cite{mischenko2020examining}.
Firestone and Scholl, in \cite{firestone2016cognition}, also note that some experimental designs in traditional `motivated perception' type studies make the studies less convincing in support of `top-down' effects. 
For example, some of these studies capture what participants recognize rather than examining what they actually see, which makes the study more about semantic priming than perception.
They also point out the risk of experimenter demands (participants reporting what they believe the experimenter expects them to report) and other forms of response bias. %, such that participants could have adjusted their response in accordance with assumptions they have about the purpose of the experiment. 
Additionally, researchers do not always show people the \textit{same} visual stimuli to compare their perception with and without the motivation or additional beliefs. 
The low-level visual differences between the experimental stimuli can weaken the results \cite{firestone2016cognition}.

To account for some of the concerns listed by Firestone and Scholl \cite{firestone2016cognition}, in the current experiment, we tried to detect potential experimenter demands by explicitly asking participants what they thought the study was about and conveying that their answers will not affect their compensation. 
We also used the same stimuli and compared participants' correlation estimation with or without the belief component. 
Our participants viewed scatterplots labeled with `X' and `Y', and later viewed the same scatterplots again but labeled with real-world variable pairs. 
They are the same stimuli, so there are no low-level differences, and the differences in correlation estimation we observed should be driven by their attitudes towards the labels alone. 

Future research could find stronger evidence for a perceptual root of the present bias by seeking evidence of biases on Just Noticeable Differences (JND) between pairs of correlations, which might be less potentially influenced by cognitive factors compared to the individual ratings used here \cite{rensink2010perception}. 
For example, if the bias effect is driven by participants reporting estimations on either side in the range of possible correlations due to uncertainty, such that they opt to report correlations in the upper range when they think the variables should be more strongly correlated and vice versa, then experiments that leverage JND comparisons should be far less influenced by this form of bias that does not require the production of correlation estimates. 
% Similarly, it would also be beneficial to record participant response times, which might lengthen in conditions where cognitively-driven biases are more strongly present. 

%\hlIf beliefs drive a change in participant percepts, their JNDs on the correlation value would be different depending on the belief they hold about the correlations depicted.But, if the bias in correlation estimation is an added effect via cognitive factors, their JNDs might not be be affected.
%For these type of experiments, it would also be helpful to record the amount of time participants spend looking at the scatterplots, as they can be useful for debugging pilots or understanding where participants are spending time in an experiment or how session length would change as more trials are added.}
%, and to test multiple correlation values for generalizability.}

%Future research could explore these underlying mechanisms, potentially contributing to existing dialogues around the power of 'top-down' effects on perception. Regardless the underlying mechanism, the present findings have implications for how we should analyze and interpret data.

%, providing promising future research directions in other disciplines such as computer science and data communication. 
%\newline
\vspace{2mm}

% future research directions - the visualization angle
\noindent \textbf{Mitigating Biases in Data Interpretation:} The current results cannot specify \textit{why} participants gave differing correlation estimates for the same scatterplot when it was labeled with `X' and `Y' compared to when it was labeled with real-world variables. 
We discussed some potential explanations in Section \ref{disc}, and future work can further test the mechanisms driving participants to misestimate the correlations depending on their beliefs.
Understanding these mechanisms can help us design tools to potentially mitigate these biases in correlation estimation.
For example, a visual analysis tool can monitor patterns of exploration or request objective decision criteria during analysis to mitigate biases \cite{wall2019toward}.
% Designers of visual data analysis tools can help reduce related sources of bias by monitoring patterns of exploration or requesting objective decision criteria, so that those tools can demonstrate an analytic bias to a user, and suggest ways of addressing it \cite{wall2019toward}. 
If the effect is driven by participants overweighing points that are more congruent with their beliefs, a visual analytic system might first survey a user's existing beliefs, and then highlight data points that might be under weighed to increase their visual salience. 
%For example, Wall et al \cite{wall2021left} demonstrate that explicitly highlighting data points or routes that seem under-explored given an analyst's interaction history can mitigate bias.  
Existing work has shown promising results in this area \cite{wall2021left}, and future work can test similar mitigation techniques in real-world contexts with more complex data sets and tasks. % , which may also produce more strongly-held beliefs. 
% More realistic environments would produce more reliable evaluations of bias mitigation tools.
%, which has the potential to strongly impact the current experimental design and data analysis pipeline to help us avoid a replication crisis.}
% , and researchers may even find stronger belief effects. 
%We also encourage future researchers to examine how we can visualize correlation data to be more effective, more persuasive, and less biasing. 
% For example, a system might explicitly highlight data points or routes that seem under-explored given an analyst's goals \cite{wall2021left}. 
% These interventions provides us confidence that we can design visualization tools to warn people of their bias and help them mitigate it. 
% We believe these tools will have a strong impact helping with experimental and data analysis decisions and avoiding the replication crisis. 
\newline
\vspace{-2mm}

% Add Section on Eliciting Beliefs 
\noindent \textbf{Eliciting Beliefs:} We elicited beliefs via self-reporting, a survey technique commonly used in existing work to capture people's state of mind \cite{joshi2015likert}, following a similar approach in measuring belief in correlations as introduced in \cite{xiong2019illusion}.
We intentionally asked participants to report their beliefs on a verbal scale instead of the same numerical scale they used to report their correlation estimates to reduce potential carryover effects and experimenter demand.
Future work should investigate the effects of reporting beliefs on a verbal versus numerical scale, which can lead to more refined quantitative models of belief elicitation.
% how closely the viewed scatterplot matches the expected correlation by having participants report their beliefs via the same numerical scale as their correlation estimation, which can lead to more refined quantitative models of belief updating.

%Belief elicitation remains a challenging task that warrants future investigation.
There exists a variety of methods to elicit beliefs from multiple disciplines (for a design space on these methods, see \cite{mahajan2022vibe}).
In behavioral economics, researchers often infer belief from a choice tasks where participants indicate their preferences \cite{samuelson1948consumption}, though more recent work has argued against its effectiveness \cite{manski2018survey}.
In visualization work, researchers have experimented with capturing belief through having participants manipulate the slope of a line in a chart \cite{karduni2020bayesian, kim2017explaining}, specialized elicitation visual interfaces \cite{kim2020bayesian, mantri2022how}, or using a Likert scale where participants rate their agreement with a statement \cite{xiong2019illusion}.
Griffiths and Tenenbaum \cite{griffiths2006optimal} demonstrated that we could approximate people's prior beliefs by asking them to make predictions.
More recently, Suchow, Pacer, \& Griffiths \cite{suchow2016design} demonstrated that we can also estimate people's priors through transmission chains, such as a game of telephone.
Future work should explore how our results might generalize (or not) to other belief elicitation methods, in addition to exploring and evaluating alternative methods to capture people's beliefs. 
\newline
\vspace{-2mm}

% Related to Bayesian framework and belief-updating
\noindent \textbf{Belief Updating:} Our work also has implications for how people update their beliefs. 
When participants reported their level of agreement with statements describing the correlation between two variables, when the variables were belief-triggering (gun/violent crime, air quality/environmental regulations), participants' prior and posterior beliefs did not significantly differ.
But when the variables were not belief-triggering (tree/names, rain/lawyers), there was a significant difference in participants' prior and posterior beliefs.

This may suggest that when participants hold strong existing beliefs (priors), seeing additional evidence does not change their beliefs. 
They stick to their strong priors, as it was built on a great deal of data from their past experiences. 
However, when participants do not have strong existing beliefs about how two variables could be correlated, just seeing one instance of a scatterplot depicting a correlation pushed them to update their beliefs, even though the scatterplot they saw was not designed to be persuasive.
The amount of change between prior and posterior beliefs can likely be used as a proxy to determine how much people are persuaded by the new data they saw, which will allow us to reverse engineer the correlation people see in data by comparing the deltas between their prior and posterior beliefs (see \cite{elhamdadi2022measure} for more discussions).
Future work can further examine how much seeing a correlation impacts belief, and how much stating a belief impacts subsequent correlation estimation, using a wider range of topics, belief elicitation methods, and correlation values. 
Additionally, for instances where participants viewed data depicting variables for which they do not have strong priors, future researchers can also examine how that data could influence participants' belief updating through exposure effects \cite{harrison1977mere}. 
%For example, perhaps seeing a stronger correlation could motivate people to update their beliefs more, and seeing a weaker correlation would cause people to update less. 
\newline
\vspace{-2mm}

\noindent \textbf{Generalizability:} We only tested two levels of correlation in the current experiment using one scatterplot design and elicited participants' correlation estimations through Pearson's r.
These decisions were made based on methodologies from prior work to control for the effect of these design-based factors and prevent the combinatorial explosion of having too many conditions \cite{wall2022vishikers}.
%, while still make an initial demonstration of the effect of belief in scatterplot interpretation and correlation estimation. 
Future work can extend this space of investigation and examine the generalizability of our effects by testing additional levels of correlation, different shapes of the data (e.g., varying the dispersion, the number of outliers, or the range of R-values), and various scatterplot designs (e.g., aspect ratios of the plot area or with regression lines).
% and various dataset types (e.g., with more, or more salient, outliers, or a different range of R-values). 

Furthermore, future work can test other methods of eliciting participant estimations. 
For example, instead of asking for an r-value, they could ask participants to indicate which points they think are 'important' or indicate the perceived correlation by drawing an ellipse or regression line. 
Such measures may be more or less sensitive to bias and may provide a more precise indication of the influence of outliers or other salient points on a trial-to-trial basis. 
These follow-up experiments can build towards a comprehensive model mapping each component of a scatterplot to its impact on viewer interpretation, potentially pinpointing the factors or mechanisms behind belief-driven misestimation of correlations.

\acknowledgments{
We thank the online participants who made this study possible. We also thank our reviewers and Akira Wada for their helpful feedback.
}

\newpage
%% BIBLIOGRAPHY %%
\bibliographystyle{abbrv}
\bibliography{reference}

\end{document}